# Role of Defects in the Paramagnetism of Fe-doped $Cs_2AgBiBr_6$ Double Perovskite

Volodymyr Vasylkovskyi[1], Olga Trukhina[1], Patrick Dörflinger[1], Mykola Slipchenko[2], Wolf Gero Schmidt[3], Timur Biktagirov[3], Anastasiia Kultaeva[1], Yakov Kopelevich[4], Vladimir Dyakonov[1]

[1]*Experimental Physics 6 and Würzburg-Dresden Cluster of Excellence ctd.qmat, Julius-Maximilian University of Würzburg, 97074 Würzburg, Germany*
[2]*Institute for Scintillation Materials, National Academy of Sciences of Ukraine, 61072 Kharkiv, Ukraine*
[3]*Physics Department, Paderborn University, D-33098 Paderborn, Germany*
[4]*Instituto de Física "Gleb Wataghin", Universidade Estadual de Campinas-UNICAMP, R. Sergio Buarque de Holanda 777, 13083-859 Campinas, Brazil*

**Abstract.** Transition-metal doping enables the introduction of spin functionality into halide double perovskites, while simultaneously modifying optical properties. Here, we combine controlled single-crystal growth, optical characterization, comprehensive electron paramagnetic resonance (EPR) spectroscopy, and first-principles modeling to identify the microscopic nature of Fe-related centers in Fe-doped $Cs_2AgBiBr_6$. Single crystals with nominal $Fe^{3+}$ concentrations up to 15% in the precursor stage were grown using a controlled-cooling method, yielding reproducible Fe incorporation up to 0.1% w.r.t. Bi, without secondary phases. Despite this low concentration, Fe doping introduces electronic states that influence optical absorption and photoluminescence. EPR measurements reveal an $S = 5/2$ $Fe^{3+}$-related center whose anisotropy follows the cubic-to-tetragonal phase transition below 120 K. Angular-dependent EPR resolves two configurations of this nearly axial spin center, with principal axes rotated by 90° and aligned with the *a/b* plane of the tetragonal lattice. Density-functional calculations attribute these centers to impurity-vacancy complexes, most likely $Fe_{Bi}$-$V_{Br}$, that stabilize in a basal configuration of the low-temperature phase. This approach resolves vacancy-coupled defect orientations, narrowing possible models to $Fe^{3+}$-vacancy complexes and establishing them as stable, orientation-sensitive spin probes of structural symmetry in halide double perovskites, while providing a microscopic basis for tuning their magnetic and optical responses.

**Keywords:** Metal-halide double perovskites, single-crystal growth, transition-metal doping, electron paramagnetic resonance, impurity–vacancy complexes, density-functional theory (DFT)

## 1. Introduction

Lead-free halide double perovskites have emerged as stable, nontoxic alternatives to lead-based perovskites, combining favorable optoelectronic properties with structural tunability, expanding the material platform for applications beyond photovoltaics. $Cs_2AgBiBr_6$ provides a potentially interesting system for investigating the impact of impurities on the properties of double perovskites [1]. In its pristine form, this indirect band-gap semiconductor is not competitive with its hybrid-organic-inorganic counterparts in solar cells. On the other hand, it exhibits intriguing opto-mechanical properties, complex anisotropic electron-lattice interactions, with effects such as photostriction [2]. Doping with transition metals provides a strategy to introduce localized spin-centers and tailor magnetic interactions in an otherwise diamagnetic host [3], potentially enabling spintronic and quantum functionalities, such as data storage, quantum communication, and or quantum sensing [4, 5, 6, 7].

Iron is among the most attractive dopants, owing to its high-spin $3d^5$ configuration and flexible valence chemistry. Previous studies demonstrated that Fe incorporation can modify the band structure and induce magnetic responses in perovskites [1]. In $Cs_2AgBiBr_6$, earlier electron paramagnetic resonance (EPR) work [8] showed that the $Fe^{3+}$ spin state accompanies the temperature-dependent structural phase transition, but the detailed origin of the anisotropy and its link to lattice defects remains unclear, because the microscopic origin of Fe-related centers and their relation to host-lattice symmetry has not yet been clarified. Precise control over the Fe-doping levels also remains a challenge, as in double perovskites, Fe can substitute either B, or B' cation sites in the respective octahedra with a different spin and charge state. Moreover, this may introduce local structural distortions [8]. Despite growing interest in spin-active double perovskites, the lack of atomic-scale understanding of impurity-related centers has remained a major bottleneck for rational materials design. Resolving this gap is essential for transforming halide perovskites from passive optoelectronic materials into platforms for quantum technologies.

Another limitation of previous studies has been the lack of an integrated experimental-theoretical framework capable of uniquely correlating EPR observables with specific atomic defect complexes. Here, we address this challenge by combining controlled synthesis, optical characterization, EPR spectroscopy, and first-principles modeling to clarify the structure and spin properties of Fe-related centers in $Cs_2AgBiBr_6$. The low but well-quantified Fe incorporation achieved through controlled cooling crystal growth enables reliable defect identification. Temperature- and angle-dependent EPR measurements reveal that $Fe^{3+}$ centers track the orthorhombic-to-tetragonal phase transition and exhibit well-defined in-plane local magnetic anisotropy. Through density-functional and multireference calculations, we demonstrate that these properties originate from impurity-vacancy complexes rather than isolated substitutional Fe ions. This integrated experimental-theoretical approach narrows defect configurations to a few plausible models, thereby establishing a generally applicable atomistic framework for the identification of spin centers in halide double perovskites, clarifying the origin of Fe-induced centers and their impact on the magnetic and optical behavior of $Cs_2AgBiBr_6$ double perovskite.

## 2. Results and discussion
### 2.1. Structural, compositional, and optical properties

The controlled cooling method yielded high-quality $Cs_2AgBiBr_6$:Fe single crystals with well-defined octahedral shapes and sharp facets, up to 6 mm in size. The crystal average size systematically decreased with increasing nominal Fe content (Figure 1a), suggesting that Fe ions alter the nucleation kinetics. Optical microscopy of [-1-1-1] facet (Figure S1 in Supporting Information (SI)) indicated triangular terrace patterns typical of layer-by-layer growth under near-equilibrium conditions and minimal surface stress, while SEM (Figure S2 in SI) revealed smooth [111] facets. Microscopy did not reveal any difference between Fe-doped and undoped samples or any significant morphological changes in the crystal structure, even in the most heavily doped samples, confirming that Fe incorporation does not crucially disrupt growth kinetics or crystal integrity.

Powder XRD patterns of all samples (Figure 1b) are consistent with the simulated $Cs_2AgBiBr_6$ diffraction profile [9], confirming single-phase cubic symmetry ($Fm\bar{3}m$) with no secondary reflections within the detection limit. At room temperature, $Cs_2AgBiBr_6$ consists of alternating corner-sharing $[AgBr_6]^{5-}$ and $[BiBr_6]^{3-}$ octahedra, with $Cs^+$ ions occupying the interstitial sites. The connectivity and tilting of these octahedra are known to be highly sensitive to subtle changes in bond lengths and cation radii, which can trigger local distortions and strongly influence electronic structure. Nevertheless, the Raman spectra at room temperature (Figure S3 in SI) show no measurable shifts or mode splitting with increasing Fe content, indicating that the incorporation of $Fe^{3+}$ at sub-percent levels preserves both the long-range order and the vibrational signature of the host lattice. All the above findings indicate that the Fe concentration in the crystal lattice is much lower than the initial concentration in the starting substance solution, meaning that the above structural methods reach their detection limits.

ICP-OES analysis revealed actual Fe incorporation levels between 0.02% and 0.1% relative to Bi (Table 1). This is far below the nominal doping ratios, yet sufficient to modify magnetic and optical properties markedly (as demonstrated in the following sections). The low Fe incorporation indicates limited solubility of Fe in $Cs_2AgBiBr_6$ host, which likely reflects the distinct bonding mechanisms of the two cations: $Bi^{3+}$ engages in partially covalent interactions stabilized by its $6s^2$ lone pair, whereas $Fe^{3+}$ forms predominantly ionic Fe-Br bonds, leading to poor chemical compatibility.

Table 1. ICP-OES results

| Sample | Bi content (%) | Fe content (%) | Fe content w.r.t. Bi (%) |
|---|---|---|---|
| $Cs_2AgBiBr_6$ | 19.1 | <0.0005 | - |
| $Cs_2AgBiBr_6$:Fe-5 | 19.2 | 0.0010 | 0.02 |
| $Cs_2AgBiBr_6$:Fe-10 | 19.1 | 0.0014 | 0.03 |
| $Cs_2AgBiBr_6$:Fe-15 | 19.2 | 0.0052 | 0.10 |

* The analytical error for Fe determination was below 0.0005%, which corresponds to the ICP-OES detection limit.

As illustrated in Figure 1a, Fe incorporation leads to systematic darkening and reduced transparency of the crystals (while color distribution across the crystal volume is homogenous; Figure S4 in SI). Most likely, this is related to defect formation within the lattice following Fe incorporation. Fe ions may replace $Bi^{3+}$ or $Ag^+$ in the lattice; however, concomitant point vacancies, antisites, and even dislocations could be formed due to charge compensation, bonding mechanisms, and local strain. Such defects scatter or absorb light, reducing the crystal's optical clarity. A similar effect was reported by Fuxiang Ji et al., where Ag-Bi disorders and isolated defect states formed during solvent evaporation crystal growth led to darkening in $Cs_2AgBiBr_6$ without doping [10].

Similarly, the UV-Vis absorption spectra (Figure 1c) show enhanced sub-bandgap absorption with increasing Fe content, further pointing to the involvement of defect-associated electronic transitions. The absorption edge broadens without a significant shift in the main excitonic feature, implying the creation of tail states rather than bandgap narrowing. These defect states act as deep traps for carriers, as supported by photoluminescence (PL) measurements, which reveal that Fe-doped samples show very weak or undetectable PL signals due to strong quenching effects. This trend is further confirmed by time-resolved PL, where the dominant decay time decreases, reflecting the progressively higher density of nonradiative trap states introduced by Fe doping (Figure S5 in SI).

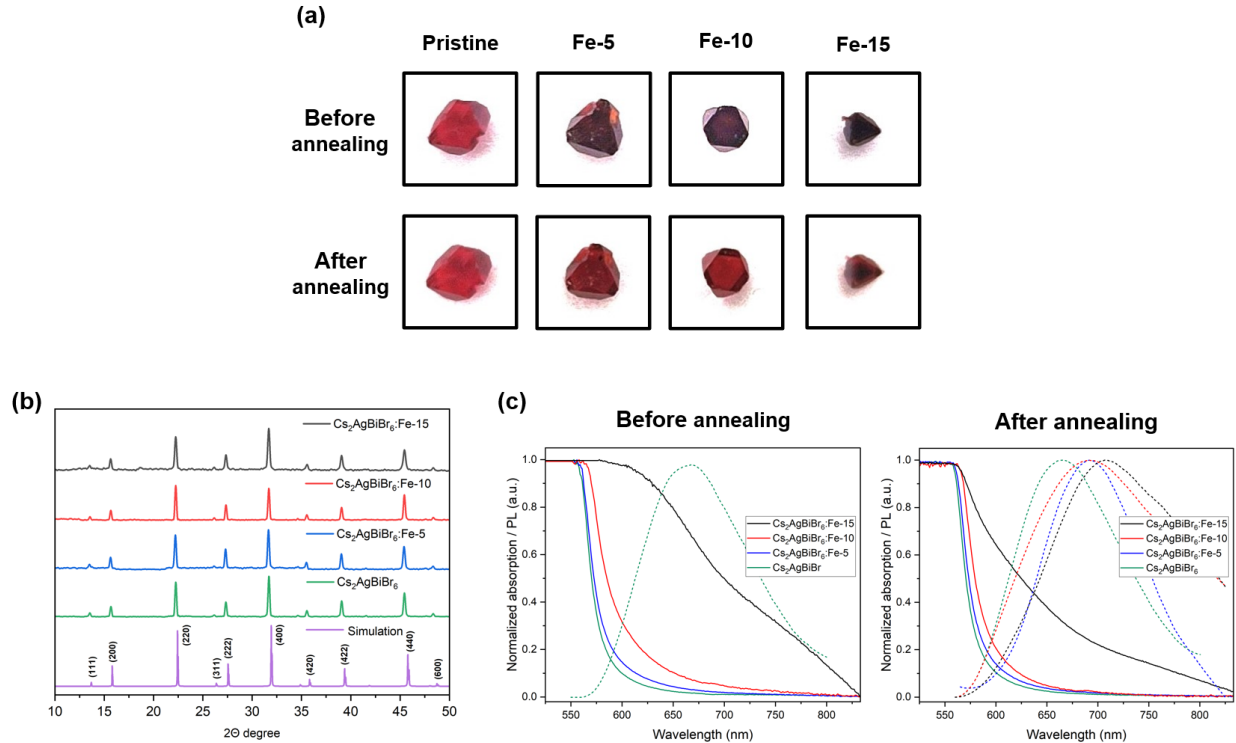

**Figure 1**. Structural and optical characterization of Fe-doped $Cs_2AgBiBr_6$. (a) Images of $Cs_2AgBiBr_6$ single crystals doped with various initial $Fe^{3+}$ concentrations before and after annealing. (b) Simulated XRD pattern, powder XRD patterns of pristine and Fe-doped $Cs_2AgBiBr_6$ single crystals. (c) Absorption (solid lines) and PL (dashed lines) spectra of non-annealed and annealed Fe-doped crystals. Before annealing, Fe-doped samples exhibited very weak or undetectable PL signals, therefore, only the PL spectrum of the pristine $Cs_2AgBiBr_6$ crystal is shown.

After thermal annealing at 300 °C in air for 30 min, which is known to reduce structural disorder in the $Cs_2AgBiBr_6$ lattice [11], optical transparency improves (Figure 1a) and sub-gap absorption diminishes (Figure 1c). Likewise, the PL spectra recover strong near-band-edge emission comparable to undoped $Cs_2AgBiBr_6$, suggesting a decrease in the density of defect states that previously facilitated nonradiative recombination (Figure 1c).

A notable observation from the PL measurements after annealing is the shift of the emission band from 665 nm (in undoped $Cs_2AgBiBr_6$) to ~690 nm in Fe-5 / Fe-10 and up to 707 nm in Fe-15 crystal (cf. Figure 1c, right panel). To rationalize this, we notice that the incorporation of $Fe^{3+}$ into, for example, the $Bi^{3+}$ site of the lattice should lead to the introduction of localized electronic states within the bandgap. These electronic states, along with those associated with remaining point defects, are expected to contribute to emission red shifts [12, 13, 14].

Optical characterization establishes a link between Fe doping and lattice disorder associated with induced defects, while also highlighting the sensitivity of these defects to thermally activated migration.

The demonstrated ability to partially reduce the defect density renders annealing as an effective strategy for suppressing defect-mediated recombination and restoring near-band-edge emission in Fe-doped $Cs_2AgBiBr_6$.

## 2.2. Electron paramagnetic resonance spectroscopy

While optical spectroscopy captures the collective impact of defect states on electronic properties and, when time-resolved, on carrier dynamics, it does not provide information about their microscopic origin. EPR spectroscopy, in contrast, provides a direct access to the local electronic structure and symmetry of spin centers. To clarify the nature of the Fe-related defects, we therefore performed temperature- and angle-dependent EPR measurements on Fe-doped $Cs_2AgBiBr_6$ single crystals.

EPR spectra of the Fe-15 samples at temperatures from 120 K to 10 K show a broad resonance dominance above 100 K, close to the free-electron $g$-factor (around $g \approx 2.00$) (Figure S6 from SI). Upon cooling below 80 K, the signal intensity increases, and the line progressively splits into a structured pattern characteristic to a high-spin system exhibiting local magnetic anisotropy that gives rise to zero-field splitting (ZFS) between $|m_S\rangle$ spin sublevels. The observed evolution coincides with the structural phase transition from orthorhombic (high-temperature) to tetragonal or monoclinic (low-temperature) symmetry, documented by neutron powder diffraction, Raman spectroscopy, and previous EPR studies [8, 15].

Importantly, in nominally 10% Fe-doped crystals, the intensity of the characteristic EPR lines is substantially lower than in the Fe-15 samples. This reduction does not reflect weaker intrinsic spin resonance but rather the markedly smaller final concentration of paramagnetic Fe-related centers incorporated into the lattice. As a result, the central broad line at $g \approx 2$ becomes disproportionately prominent, which does not originate from the sample but arises from the quartz crystal holder (Figure S7 from SI), and becomes visible only because the main sample-related transitions are strongly suppressed at this lower effective doping level. As shown in Figure 2a, the EPR response becomes progressively more structured with increasing Fe content, indicating that higher doping levels produce a larger amount of Fe-related defect complexes. The simulation of the EPR spectrum acquired at the 40 K mark suggests that it originates from an $S = 5/2$ spin system, consistent with $Fe^{3+}$ in a well-defined axially symmetric coordination environment.

As shown in Figure 2b, thermal annealing drastically reduces the intensity of this $S = 5/2$ EPR signal. In light of the annealing behavior of the defect-related optical signatures discussed above, this observation implies that the spin center should not be associated with a simple individual $Fe^{3+}$ ion, but rather with $Fe^{3+}$ coupled with another defect (e.g., a halide vacancy).

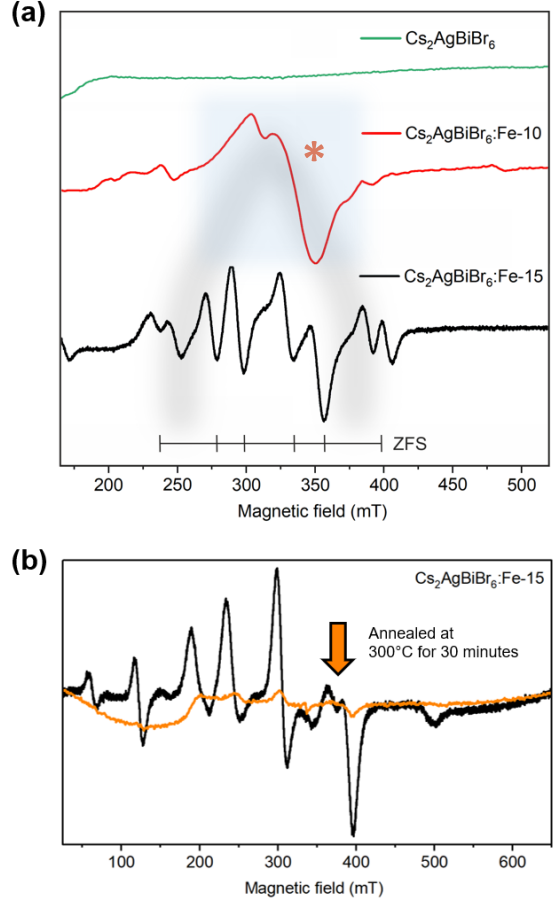

**Figure 2.** (a) EPR spectra of undoped and Fe-doped Cs$_2$AgBiBr$_6$ crystals. The wide line on the spectrum for Fe-10, which corresponds to the signal from the quartz crystal holder, is marked with an asterisk and highlighted in blue. (b) EPR spectra of Cs$_2$AgBiBr$_6$:Fe-15 crystal before and after annealing.

To determine the orientation and symmetry of this spin center, EPR spectra of the Fe-15 sample were recorded at 40 K (in the tetragonal *I*4/*m* phase) with the crystal gradually rotated in the external magnetic field (Figure 3). The angular dependence of the resonance field for an axially symmetric $S = 5/2$ ion follows directly from the spin Hamiltonian

$$\hat{H} = \mu_B \mathbf{B} \cdot \mathbf{g} \cdot \hat{S} + D\left[\hat{S}_z^2 - \frac{1}{3}S(S+1)\right] + E(\hat{S}_x^2 - \hat{S}_y^2),$$

where the first term describes the Zeeman interaction and the second and third terms represent the axial and rhombic (non-axial) ZFS, respectively. In the axially symmetric case, $E \approx 0$, and the principal values of the g factor are $g_z = g_\parallel$, and $g_x = g_y = g_\perp$. An illustration of the angular dependence for a purely axial $S = 5/2$ center with isotropic g is provided in Figure 3a. In this case, when the field $B$ forms an angle $\theta$ with the symmetry axis z, the resonance conditions for the $\Delta m_S = \pm 1$ transitions are given to the first order by $h\nu \approx g\mu_B B + D'(\theta)(m_S - \frac{1}{2})$, where the angular dependence of the ZFS is simply $D'(\theta) =$ $D(3\cos^2\theta - 1)$. The spacing between the EPR lines is the largest when $\theta = 0$, i.e., when the magnetic field direction coincides with the center's symmetry axis. The spacing collapses when $\theta$ approaches 54.7°, the magic angle where $3\cos^2\theta = 1$ (cf. Figure 4a).

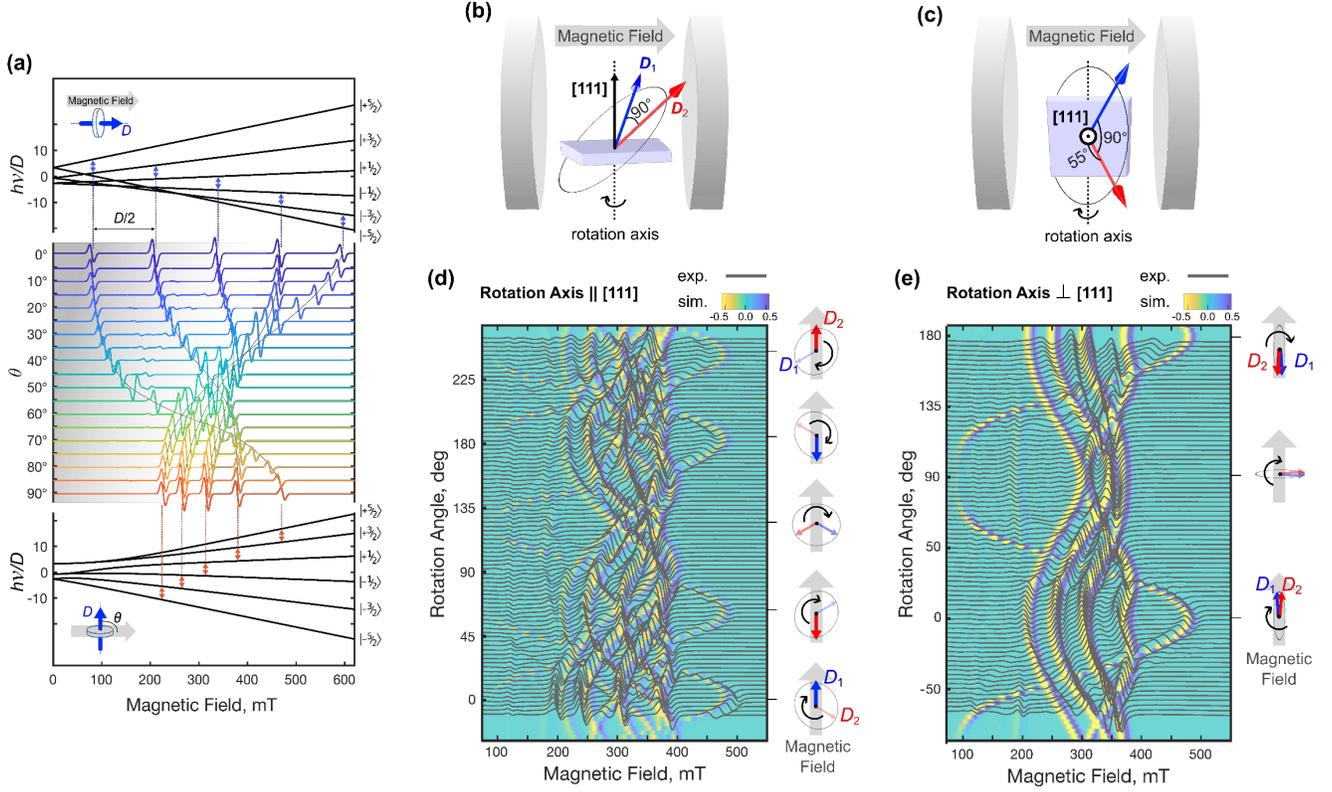

**Figure 3.** Angular dependent EPR spectra of the $Cs_2AgBiBr_6$:Fe-15 crystal measured at 40 K when the crystal was rotated in the external magnetic field. (a) Example of energy level diagrams and allowed $\Delta m_S = \pm 1$ transitions of an arbitrary axially symmetric S = 5/2 spin center with external magnetic field direction parallel (top panel) and perpendicular (bottom panel) to the symmetry axis. The corresponding angular dependence of the simulated EPR transitions is shown in the middle panel. (b, c) Two orientations of the sample with respect to the rotation axis used in the experiment. (d, e) The corresponding angular dependences measured for the setups depicted in (b) and (c). The yellow/purple contours represent the maxima/minima of the simulated EPR spectra assuming two $S = 5/2$ spin centers with equivalent spin Hamiltonian parameters but with the symmetry axes (denoted as $D_1$ and $D_2$) related by 90°. The directions of $D_1$ and $D_2$ with respect to the rotation axis and magnetic field are sketched in (d, e) for selected crystal orientations as well as in (b) and (c).

In the experiment, we used two rotation planes, schematically shown in Figure 3b, c. In the first configuration, the external magnetic field was rotated in the plane parallel to the largest crystal facet, most likely normal to [111]; in the second, the rotation axis was turned by 90°, effectively sampling two orthogonal crystallographic directions. For each configuration, spectra were collected in 3° steps, producing a two-dimensional angular map of resonance-field positions (Figure 3d, e). The result is strikingly regular: the pattern consists of several branches that shift periodically with the rotation angle, characteristic of a high-spin S = 5/2 center with nearly axial symmetry.

Numerical simulation using the above spin Hamiltonian (depicted as the color map superimposed with the experimental EPR spectra in Figure 3d, e) reproduces the measured angular patterns for parameters $D = 2$ GHz, $E \approx 0$, $g_\parallel = 2.08$, and $g_\perp = 2.02$. However, a single orientation fails to describe both datasets. The best fit requires two equivalent centers, denoted $D_1$ and $D_2$, with identical spin-Hamiltonian parameters but with their local z axes rotated by 90° relative to each other. This configuration perfectly accounts for the periodicity and relative phase of the two angular dependences shown in Figure 3d, e. Like the textbook case of Figure 3a, both angular dependences exhibit the largest splitting when the external magnetic field aligns with the projection of either $D_1$ or $D_2$ symmetry axis onto the rotation plane. In the first angular dependence map (Figure 3b, d), the projections of $D_1$ and $D_2$ are offset by 60°, while in the second crystal orientation (Figure 3c, e) their projections coincide throughout the entire experiment, thus explaining the observed 180° periodicity.

The simulation results indicate that both $D_1$ and $D_2$ axes are oriented by about 55° with respect to the [111] axis of the single crystal. In structural terms, it implies that the local axial symmetry of the $Fe^{3+}$ centers is most likely aligned along the in-plane *a* and *b* directions of the tetragonal lattice. The equivalence of $D_1$ and $D_2$ therefore reflects the presence of two symmetry-related variants of the same defect, stabilized along orthogonal tetragonal domains within the basal plane. Interestingly, no analogous signal corresponding to a center aligned along the c direction is detected. Otherwise, this c-axis-oriented $Fe^{3+}$ center would manifest itself as a resolved six-line pattern at the 120° mark in Figure 3d and at 90° in Figure 3e, i.e., when B matches its z direction.

The absence of a third orientation of the $Fe^{3+}$ center is particularly striking. In the high-temperature cubic phase of $Cs_2AgBiBr_6$, all three ⟨100⟩ directions are crystallographically equivalent, and one would therefore expect up to three magnetically equivalent Fe sites aligned along a, b, and c. Upon cooling below 120 K, following the cubic-to-tetragonal transition, this degeneracy is lifted as the lattice elongates along one axis and the [111]-type octahedral framework undergoes further distortion under 80 K (Figure S6). If the Fe-related center simply followed the lattice symmetry, one might still anticipate an additional variant oriented along the tetragonal c direction. Its absence in the low-temperature EPR spectra implies that the corresponding configuration is either energetically unfavorable or possesses a markedly different ZFS that shifts it outside the X-band detection range. An alternative, purely structural explanation could be the formation of twin domains rotated by 90°, which would yield two sets of in-plane axes but no distinct c-axis component. While this possibility cannot be ruled out, the optical and annealing data suggest a defect-driven rather than domain-driven origin. To test this hypothesis, we turned to first-principles modeling, examining whether two distinct but symmetry-related Fe-defect complexes within a single tetragonal domain can reproduce the observed in-plane anisotropy.

## 2.3. DFT calculations

To rationalize the microscopic origin of the $Fe^{3+}$ center revealed by EPR, we performed first-principles modeling. All DFT calculations were carried out for the tetragonal $I4/m$ phase of $Cs_2AgBiBr_6$ using a supercell approach with periodic boundary conditions. Our aim was to assess the relative thermodynamic stability of possible Fe-related configurations and evaluate their ZFS tensors.

The most natural starting point is isovalent substitution of $Bi^{3+}$ by $Fe^{3+}$, denoted $Fe_{Bi}$ and shown in Figure 4a. This configuration maintains overall charge neutrality and preserves the local octahedral environment of Bi. However, structural relaxation reveals that the Fe ion does not remain exactly on the ideal Bi site: small off-center displacements of 0.02-0.10 Å are energetically favorable, either along $c$ or within the basal plane, with slight energy differences below 10 meV. The shallow potential energy landscape reflects that $Fe^{3+}$ cannot fully engage in a bonding network formed by the stereochemically active $Bi^{3+}$ $6s^2$ lone pairs. In contrast to Bi, the $d^5$ $Fe^{3+}$ ion forms more ionic Fe-Br bonds, leading to several nearly degenerate off-center minima rather than a single well-defined localization site.

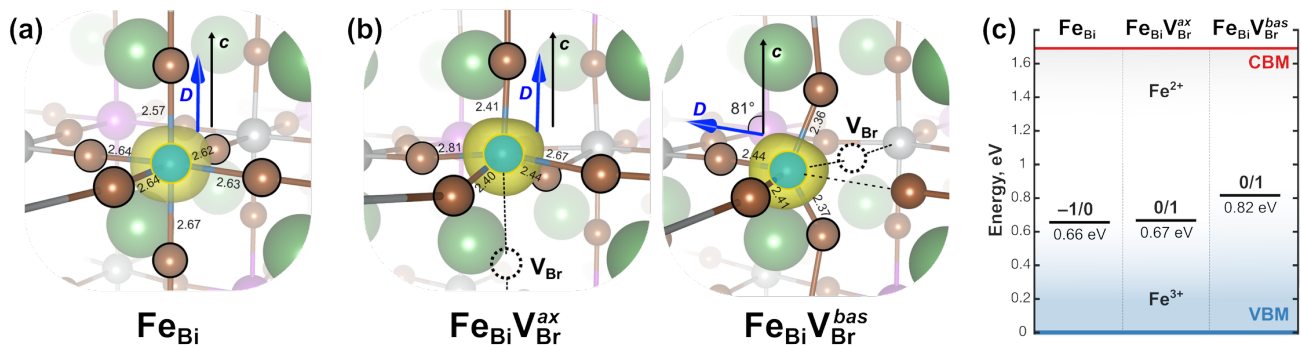

**Figure 4.** DFT optimized structures of $Fe^{3+}$ defect models considered in this work: (a) $Fe_{Bi}$ in its neutral charge state, (b) positively charged axial and basal complexes of $Fe_{Bi}$ and $V_{Br}^+$. Yellow blobs illustrate the spin density distribution. (c) Calculated charge-transition levels of $Fe_{Bi}$, $Fe_{Bi}V_{Br}^{ax}$, and $Fe_{Bi}V_{Br}^{bax}$.

To assess the associated magnetic anisotropy and its direction, we first computed the spin-spin contribution (caused by magnetic dipole-dipole coupling) to the ZFS tensor of $Fe_{Bi}$ [16]. The principal component $D_{zz}$ is oriented along $c$ and corresponds to a nearly axial tensor ($E/D \approx 0.01$). The resulting $D$ value, 103 MHz, is more than an order of magnitude smaller than the experimental value of 2 GHz.

It should be noted, however, that the quantitative prediction of ZFS parameters for high-spin $3d^5$ systems such as $Fe^{3+}$ and $Mn^{2+}$ is extremely challenging [17, 18]. In general, the spin-spin part alone accounts only for a fraction of ZFS. At the same time, the second – often dominant – contribution caused by spin-orbit coupling treated within a single-determinant picture of DFT is known to yield unreliable ZFS values for $Fe^{3+}$ since it neglects near-degeneracies and mixing of spin-orbit states.

Multireference approaches, such as CASSCF, are required for quantitative accuracy but can only be applied to finite clusters rather than extended periodic systems. We therefore combine both perspectives by complementing our supercell-based spin-spin ZFS calculation (which gives a reliable anisotropy direction) with a multireference calculation using a cluster model cut from the DFT-optimized supercell. CASSCF calculation on a resulting $[FeBr_6]^{3-}$ cluster confirmed a small magnetic anisotropy ($D \approx 0.4$ GHz). We thus rule out isolated $Fe_{Bi}$ as the origin of the observed EPR center.

Next, we consider that $Fe_{Bi}$ might form a complex with a native defect, which should result in increased ZFS. The most plausible candidate is bromine vacancy, $V_{Br}$. Such $V_{Br}$ defects are known to be abundant in $Cs_2AgBiBr_6$, with their 1+ charge state being dominant [19]. The presence of $V_{Br}^{1+}$ in the direct vicinity of $Fe_{Bi}$ would naturally lift the local quasi-octahedral symmetry of $Fe_{Bi}$, enhancing the magnetic anisotropy, while ensuring that the Fe ion retains its 3+ charge state and $S = 5/2$ configuration. This model also aligns with the observed correlation of the EPR signal with optical defect signatures as well as its annealing behavior.

Two distinct geometries of the $Fe_{Bi} - V_{Br}^+$ complex were identified (cf. Figure 4b): an *axial* complex denoted as $Fe_{Bi}V_{Br}^{ax}$, where $V_{Br}^+$ occupies a nearest-neighbor apical Br site along *c*, and a *basal* complex $Fe_{Bi}V_{Br}^{bas}$, where the vacancy resides in the *ab* plane. The axial defect produces a distorted square-pyramidal $FeBr_5$ coordination. In contrast, the basal one yields a quasi-tetrahedral $FeBr_4$ environment with two equatorial and two apical ligands, leaving one equatorial Br virtually uncoordinated. This directly discriminates between axial and basal anisotropy directions, which is precisely the scenario expected from single-crystal EPR.

Energetically, the basal configuration is favored by $\Delta E = 22$ meV. Even such a modest energy difference translates into a strong population bias: the Boltzmann factor $exp(\Delta E/kT)$ implies that at 122 K, when the phase transition begins, the basal complex is $\approx 100$ times more abundant than the axial one. If the structural transformation persists to lower temperatures - as implied by the evolution of the EPR spectrum in Figure 4a - then at 40 K, where EPR spectra were recorded, the ratio would exceed 500.

Subsequently, we computed the binding energy of the $Fe_{Bi}V_{Br}^{bas}$ defect pair: $E_{bind} = E[Fe_{Bi}V_{Br}^{bas}] + E_{bulk} - (E[Fe_{Bi}] + E[V_{Br}^+])$, where $E[\cdot]$ are the total energies of DFT-optimized supercells containing individual defects and $E_{bulk}$ is that of pristine $Cs_2AgBiBr_6$. The resulting value of –2.01 eV indicates strong binding and further underscores the favorability of this complex.

We also computed the charge-transition levels of $Fe_{Bi}V_{Br}^{bas}$ and $Fe_{Bi}V_{Br}^{ax}$ in comparison with isolated $Fe_{Bi}$. As shown in Figure 4c, all configurations favor the $S = 5/2$ $Fe^{3+}$ state for Fermi levels located in the lower and mid-gap regions. Previous resistivity measurements place the Fermi level of intrinsic $Cs_2AgBiBr_6$ single crystals near 0.79 eV above the VBM [20], i.e., close to mid-gap and slightly below the calculated $Fe^{3+}/Fe^{2+}$ transition of the $Fe_{Bi}V_{Br}^{bas}$ complex. In our crystals, we assume a similar intrinsic Fermi level, which stabilizes $Fe^{3+}$ as the dominant charge state.

The principal direction of the computed spin-spin ZFS for the $Fe_{Bi}V_{Br}^{bas}$ complex is almost perpendicular to the *c*-axis (81°). The principal value is 0.3 GHz, still smaller than experiment but expectedly larger than for isolated $Fe_{Bi}$. To capture the spin-orbit contribution missing in periodic DFT, we again modeled a finite $[FeBr_4]^{1-}$ cluster corresponding isolated from the DFT-optimized supercell. Multireference CASSCF calculations yield $D \approx 5.9$ GHz, a reasonable match to the experimental 2 GHz value, considering the crude approximation. The principal $D$ axis lies within the plane of the missing Br ligand, reproducing the in-plane orientation of the anisotropy tensor inferred from EPR.

Altogether, the calculations establish that $Fe^{3+}$ at the Br site in $Cs_2AgBiBr_6$ tends to form a defect complex with a neighboring positively charged bromine vacancy. The energetics of the $Fe_{Bi}V_{Br}$ complexes explain the dominance of in-plane $Fe^{3+}$ centers over the axial ones observed experimentally and their persistence below the phase-transition temperature. This model thus bridges the EPR, optical, and structural observations and highlights that the $Fe^{3+}$–vacancy pairs can be used as spin probes to determine the direction of the tetragonal axis in the low-temperature phase.

**Conclusions**

In conclusion, we successfully synthesized Fe-doped $Cs_2AgBiBr_6$ single crystals with various Fe concentrations and performed a comprehensive analysis of their structural, optical, and spin properties. Optical spectroscopy revealed that Fe doping introduces defect states whose density can be tuned and partially healed by thermal annealing.

Combining single-crystal EPR spectroscopy with first-principles modeling, we have identified the microscopic origin of the introduced Fe-related $S = 5/2$ spin center and elucidated its response to the structural transformation of the host lattice. We show that the $Fe^{3+}$ signal arises not from isolated substitutional ions but from defect complexes, most likely $Fe_{Bi}$-$V_{Br}^+$. Remarkably, our results suggest that these complexes stabilize in the *basal* plane of the tetragonal phase, while the *axial* configuration is unfavorable. As a result, these spin centers exhibit two orthogonal orientations within the *a*/*b* plane of the low-temperature lattice, which renders them as promising EPR probes of lattice symmetry.

These findings highlight the complex interplay between dopants, lattice defects, and structural phase behavior in $Cs_2AgBiBr_6$. They show that controlled transition-metal doping in halide double perovskites can create reproducible, orientation-defined paramagnetic centers rather than disordered magnetic phases, offering a route to deliberately engineer spin-active centers in these chemically stable, lead-free materials.

**Materials and methods**

**$Cs_2AgBiBr_6$:$Fe^{3+}$ crystal growth**

The growth of $Cs_2AgBiBr_6$ perovskite single crystals with varying initial $Fe^{3+}$-doping ratios was performed using a controlled cooling crystallization growth technique. This method minimizes the formation of secondary phases and enables the growth of single crystals up to 1 cm in size without using seed crystals [21]. The materials used for crystal growth included: Cesium bromide (CsBr, 99.999%) and Ferric bromide ($FeBr_3$, 98%) from "Sigma-Aldrich"; Bismuth tribromide ($BiBr_3$, 98%) and silver bromide (AgBr, 99%) from "Alfa". Hydrobromic acid (HBr, 48%) from "Acros Organics" was used as both the solvent and a crystal growth medium, facilitating controlled chemical reactions necessary for crystal formation. All reagents were used without further purification. To prevent oxidation and contamination, all reagents were handled and precisely weighed in a nitrogen-filled glovebox with water and oxygen content lower than 2 ppm. Solid CsBr, $BiBr_3$, AgBr, and $FeBr_3$ were dissolved in 10ml of HBr in a glass flask. The $FeBr_3$ concentration was adjusted relative to $BiBr_3$ to achieve varying initial molar concentrations of $Fe^{3+}$ doping (0%, 5%, 10%, 15%).

The flask containing the precursor solution was placed in a silicon oil bath located on the hot plate. The oil bath was heated to 110 °C and maintained at 110 °C for 5 hours to ensure complete dissolution of the reagents. Afterward, the temperature was gradually reduced to room temperature at the rate of 1°C per hour to precisely control nucleation and crystal growth. The resulting single crystals were harvested, washed with dichloromethane and isopropanol to remove precursors from their surface, and then dried under ambient conditions.

**Characterization of Fe-doped $Cs_2AgBiBr_6$ single crystals**

X-ray diffraction (XRD) measurements were performed using a General Electric XRD 3003 TT system with a monochromatic copper (Cu) Kα radiation source to determine the crystalline structure. For XRD analysis, the crystalline samples were ground into powders and spread onto a polymer sample holder. Measurements were performed in Bragg-Brentano geometry within the 2Theta range from 10° to 50° with a step size of 0.005° and a dwell time of 5 seconds per step, at room temperature.

For inductively coupled plasma optical emission spectroscopy (ICP-OES) analysis, the crystal samples were ground into powders, and a microwave-assisted acid digestion was then performed using nitric acid ($HNO_3$) and hydrogen peroxide ($H_2O_2$) to dissolve the powder samples. Despite efforts to fully dissolve the samples, the dissolution was incomplete, leaving a yellow precipitate. However, the precipitate did not contain Fe or Bi, which was confirmed by Energy-dispersive X-ray (EDX) analysis. Bismuth and iron concentrations in the solutions were quantified by ICP-OES, calibrated with freshly prepared external calibration standards derived from certified reference solutions.

The UV-VIS absorption measurements were performed using a Jasco V-630 UV-vis spectrometer. Photoluminescent (PL) spectroscopy and transient photoluminescent (trPL) measurements were carried out using an FLS 980 spectrometer from "Edinburgh Instruments Ltd" with a 375 nm excitation wavelength.

Electron paramagnetic resonance (EPR) measurements were performed using a Magnettech MiniScope MS 5000 EPR spectrometer equipped with an Oxford ESR 900 He flow cryostat. The measurements were conducted over the temperature range of 300 K to 10 K, operating at a frequency of 9.42 GHz and microwave power of 1 mW. For angular-dependent measurements, the sample was mounted on a holder with one axis aligned perpendicular to the magnetic field, and a goniometer was employed to achieve precise rotational measurements. The simulations of the powder EPR spectra were carried out using the Matlab toolbox EasySpin [22].

**Computational**

All calculations were performed for the low-temperature I4/m structure in a 2×2×2 supercell (160 atoms using the Quantum ESPRESSO software package [23, 24].

PBEsol exchange-correlation functional [25] was employed for structural relaxation, and hybrid HSE06 functional [26] was used for electronic structure calculations. Spin-spin ZFS tensors were computed using periodic DFT (as in Ref. [27]). To estimate spin-orbit contributions to the ZFS tensor, additional complete active space self-consisted field (CASSCF) calculations were carried out using the ORCA software package v.6.0 [28] for cluster models cut from optimized supercells. The active space consisted of five electrons distributed in the five metal-based 3d orbitals (CAS(5,5)), with the state-average CASSCF calculation incorporating one sextet, 24 quartet, and 75 doublet roots.

**Acknowledgments**

V.V., A.K., V.D. acknowledge financial support by the Deutsche Forschungsgemeinschaft (DFG, German Research Foundation) through the Würzburg-Dresden Cluster of Excellence ctd.qmat – Complexity, Topology and Dynamics in Quantum Matter (EXC 2147, project-id 390858490). Y.K. was financially supported by FAPESP and CNPq. Authors also thank Stephan Braxmeier from Center for Applied Energy Research (CAE) for SEM and EDX measurements and SGS INSTITUT FRESENIUS GmbH for the ICP-OES Analysis.

**Declarations**

**Conflict of interest -** The authors declare that they have no conflict of interest.

**Data Availability**

The data obtained and/or analyzed during the study are not publicly available but are available from the corresponding author upon reasonable request.